\documentclass[12pt]{article}

\usepackage[utf8]{inputenc}
\usepackage[T1]{fontenc}
\usepackage{amsmath}
\usepackage{url}
\usepackage{color}
\usepackage{graphicx,subfig}
\usepackage[margin=2cm]{geometry}

\usepackage{textcomp }
\usepackage{lineno}
\modulolinenumbers[5]
\usepackage{multirow}
\usepackage{float}
\usepackage[superscript,biblabel]{cite}

\usepackage{xr}
\usepackage{braket}
\usepackage{scicite}

\usepackage{times}
\newenvironment{sciabstract}{%
\begin{quote} \bf}
{\end{quote}}

\begin{document}
\title{Imaging local diffusion in microstructures using NV-based pulsed field gradient NMR}

\author{\hspace{-2cm}Short title: NV-based diffusion imaging in microstructures\\\hspace{-.7cm}Fleming Bruckmaier$^1$, Robin D. Allert$^1$, Nick R. Neuling$^1$, Philipp Amrein$^2$,\\\hspace{-2cm} Sebastian Littin$^2$, Karl D. Briegel$^1$, Philip Sch\"{a}tzle$^3$,\\\hspace{-2cm} Peter Knittel$^4$, Maxim Zaitsev$^2$, Dominik B. Bucher$^{1,5\,\ast}$
\\
\hspace{-2cm}\normalsize{$^1$Department of Chemistry, TUM School of Natural Sciences, Technical University of Munich, Germany}\\
\hspace{-2cm}\normalsize{$^2$Department of Radiology, University Medical Center Freiburg, Germany}\\
\hspace{-2cm}\normalsize{$^3$University of Freiburg, Department of Sustainable Systems Engineering (INATECH), Emmy-Noether-Str. 2, 79110}\\
\hspace{-2cm}\normalsize{$^4$Fraunhofer Institute for Applied Solid State Physics, Tullastr. 72, 79108 Freiburg, Germany}\\
\hspace{-2cm}\normalsize{$^5$Munich Center for Quantum Science and Technology (MCQST), Schellingstr. 4, 80799 München, Germany}\\
\hspace{-2cm}\normalsize{$^\ast$Corresponding author: Prof. Dr. Dominik Bucher; E-mail: Dominik.Bucher@tum.de }
}

\date{}

\renewcommand\thesubsection{\Alph{subsection}}
\renewcommand\thesubsubsection{\thesubsection.\Roman{subsection}}


\baselineskip24pt


\maketitle 

\begin{sciabstract}
Understanding diffusion in microstructures plays a crucial role in many scientific fields, including neuroscience, cancer or energy research. While magnetic resonance (MR) methods are the gold standard for diffusion measurements, spatial encoding in MR imaging has limitations. Here, we introduce nitrogen-vacancy (NV) center based nuclear magnetic resonance (NMR) spectroscopy as a powerful tool to probe diffusion with an optical readouts. We have developed an experimental scheme combining pulsed gradient spin echo (PGSE) with optically detected NV-NMR spectroscopy, which allows for the local quantification of molecular diffusion and flow within microscopic sample volumes. We demonstrate correlated optical imaging with spatially resolved PGSE NV-NMR experiments probing anisotropic water diffusion within a model microstructure. Our optically detected PGSE NV-NMR technique opens up prospects for extending the current capabilities of investigating diffusion processes with the future potential of probing single cells, tissue microstructures, or ion mobility in thin film materials for battery applications.

\end{sciabstract}

\flushbottom

\thispagestyle{empty}

\section*{Introduction}
Molecular and ion diffusion plays a major role in many aspects of physics, chemistry, and biology, ranging from nutrient transport in organisms \cite{spector2009,Gulani2001}, pattern formation \cite{Landge2020}, to the reactivity in chemical reactions \cite{noyes1956} or the functioning of modern batteries \cite{Yi2020}. Nuclear magnetic resonance (NMR) spectroscopy is one of the prevalent methods for probing diffusion \cite{Callaghan2011,neil1997} which was first described in 1965 by Stejskal and Tanner \cite{Stejskal1965}. Since then, the technique has developed rapidly and is nowadays used on a daily basis in the form of diffusion weighted magnetic resonance imaging in medicine \cite{Fieremans2018,Johnson2023,jones2010diffusion,LeBihan2003,Lundell2021,Novikov2018}.
However, magnetic resonance methods are limited by the low net nuclear magnetization of the sample, which often leads to the signal-to-noise ratio (SNR) to constrain widespread use of this otherwise powerful technology. Moreover, the spatial resolution in liquid-state magnetic resonance imaging (MRI) techniques is limited by the molecular diffusion which reduces the localisation imposed by the applied magnetic field gradient encoding \cite{Callaghan1988}. Also the intrinsic diffusion weighting, that is imposed on the sample by the imaging gradients and spoiling gradient pulses themselves, may present challenges in some studies \cite{Lasi2017} For the above-mentioned reasons, assessing diffusion with micrometer resolution within thin film materials, biological tissue or even for single cells remains extremely challenging for the NMR methodology.\\
An elegant solution to overcome these problems is the nitrogen vacancy (NV) center in diamond which is an atom-sized quantum sensor for magnetic fields \cite{Balasubramanian2008,Maze2008}. Due to its spin state-dependent fluorescence, optically detected magnetic resonance (ODMR) experiments can be performed spatially resolved in two dimensions which translate the local magnetic field into an optical signal. NV centers have been used to conduct NMR experiments on unprecedented length scales \cite{Mller2014,Sushkov2014,Staudacher561,Mamin2013,liu2021surface} and allow the detection of high spectral resolution NMR signals from microscopic sample volumes \cite{Glenn2018,Arunkumar2021,Smitseaaw7895,Bucher2020,Allert2022, Allert2022Micro}.

\begin{figure}[H]
\centering
\includegraphics[width=0.3\linewidth]{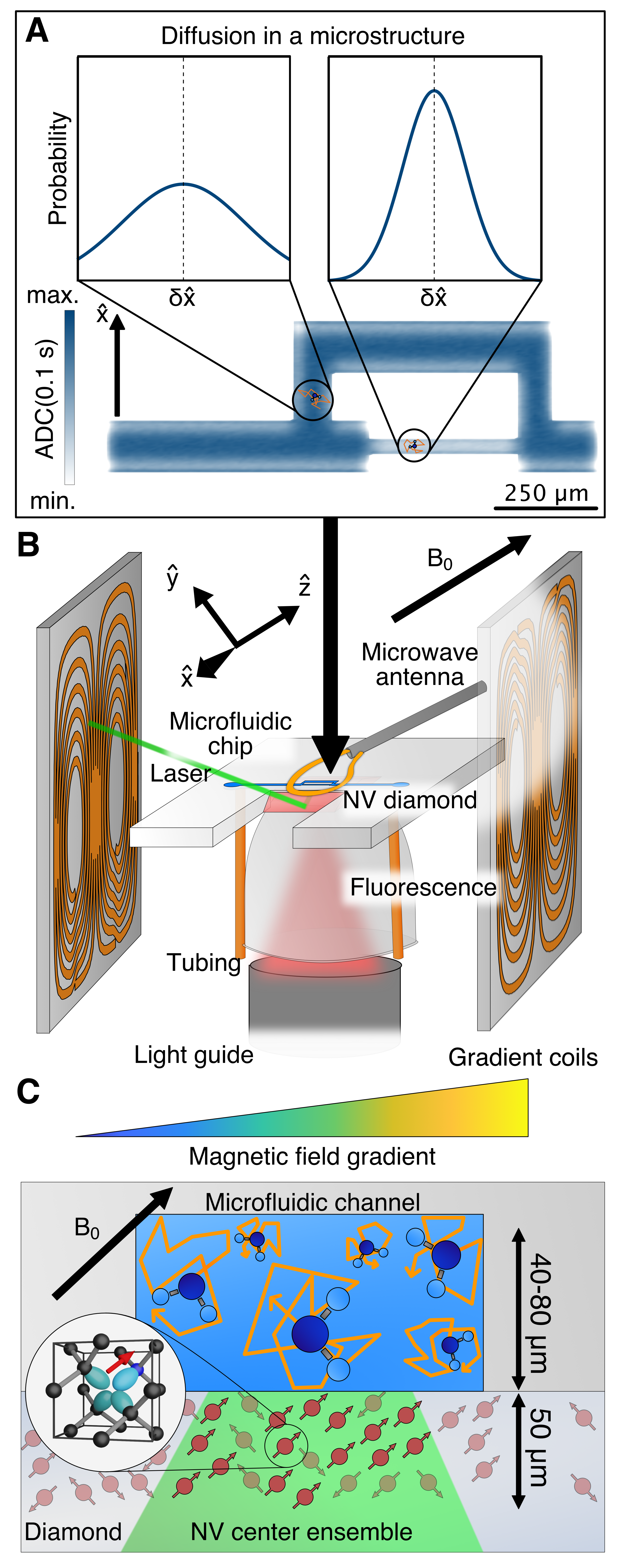}
\caption{\textbf{Principles of NV-based diffusion imaging within microstructures.} \textbf{A)} Conceptual schematic of diffusion within a microstructure. The apparent diffusion coefficient (ADC) at the two marked locations differs strongly in the $\hat{x}$ direction, since the free diffusion length is on the same scale as the microstructure itself. The probability to find a diffusing particle at a distance $\delta \hat{x}$ from its original position (dashed line) after diffusing for 0.1 s is displayed in the two plots on the top. The microstructure itself is color coded according to the simulated ADC. \textbf{B)} Experimental setup. A diamond chip (red) with a highly dense surface doped NV layer is glued into the microfluidic chip (light grey) and placed in between three pairs of magnetic field gradient coils. Each pair produces a $\hat{B}_0$ gradient along one of the cardinal directions $\hat{x}$, $\hat{y}$ and $\hat{z}$. The whole experiment is imaged using an optical microscope from above. A (green) laser enters the diamond chip and excites the NV centers in the surface layer, defining the measurement location (see also panel C). The red NV fluorescence for signal readout is collected and directed to a photodiode using a liquid light guide. The NV electronic spin, used for the quantum sensing protocol is driven by a microwave (MW) antenna on top of the microfluidic chip. \textbf{C)} The water sample is confined by a microfluidic channel, whose bottom wall is formed by the NV sensor. Water molecules interacting with the channel walls are hindered in their diffusion and will have a lower ADC. External magnetic field gradients encode the position of the water molecules and allow for the measurement of their ADCs.}
\label{fig: Setup Schematic and grad}
\end{figure}

 This technology is well suited for the investigation of diffusion phenomena on the microscopic level, due to its optical readout, high spatial resolution and capability of measuring coherent NMR signals. As a rule of thumb, the detection volume of the NV sensor corresponds to the laser spot size and the thickness of the NV layer (details on spatial contributions to the signal in NV-NMR can be found in the SM section 4). In contrast to macroscopic diffusion-based MRI experiments, the NV sensor enables the local detection of the NMR signals on a length scale similar or smaller than the average distance a water molecule will have diffused within the timescale of a typical NMR experiment. If the molecule encounters a barrier, the average displacement is reduced compared to the case of free diffusion. A microscale NV-NMR is the promising tool for probing diffusion within microstructures due to its superb localization and potentially a higher sensitivity for micro-scale sample volumes, as shown in Fig. \ref{fig: Setup Schematic and grad} A.\\
In this work, we realize microscopic imaging of molecular diffusion with NV-NMR. We first developed magnetic field gradient coils and designed pulse sequences that combine pulsed field gradients with the NV-NMR detection scheme. This allows us to perform pulsed gradient spin echo (PGSE) experiments to detect diffusion within picoliter sample volumes. In the first series of experiments, we measure water flow within a microfluidic channel. In the second step, a water-soluble polymer is added to probe its influence on water diffusivity. Finally, we demonstrate the capabilities of our technique for detecting local water diffusion within a microstructure. Spatially resolved diffusion NV-NMR measurements within a microfluidic model structure show anisotropic diffusion according to the restrictions given by the local geometry and structure.

\section*{Results}

\begin{figure}[H]
\centering
\includegraphics[width=1\linewidth]{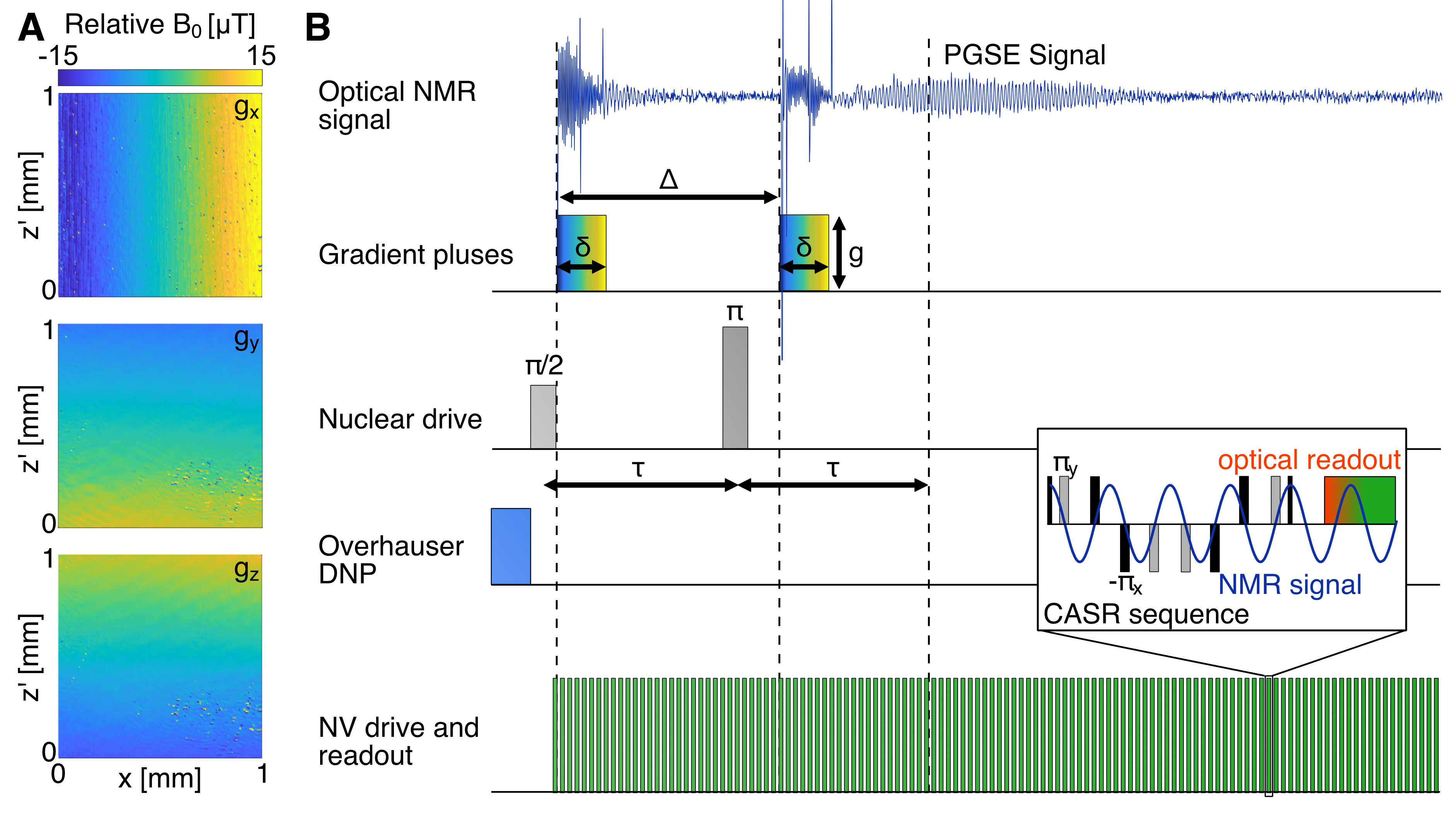}
\caption{\textbf{Principle of the pulse gradient spin echo (PGSE) NV-NMR sequence.} \textbf{A)} Magnetic field gradients: Measured $B_0$ gradients along the three cardinal directions (see Fig. \ref{fig: Setup Schematic and grad} B) using an NV wide-field magnetic imaging setup. The magnetic field gradients are measured along the diamond surface x and z' direction (parallel to the diamond surface). \textbf{B)} Measurement pulse sequence: After hyperpolarizing the sample spins using Overhauser DNP, a RF $\pi/2$-pulse at the Larmor frequency of the protons initializes the free induction decay (FID). After a time $\tau$, a $\pi$-pulse refocuses the sample nuclear spin magnetization leading to a spin echo (experimental data in blue). For the PGSE experiment, magnetic field gradient pulse with equal duration $\delta$ and strength $g$ are applied before and after the nuclear spin $\pi$-pulse (blue), separated by a total time $\Delta$. These magnetic field gradient pulses encode the position of the nuclear spins in their phase and any translation or diffusion during the time $\Delta$ will reduce the total spin echo signal. The ADC can be obtained by measuring the spin echo amplitude as a function of the applied magnetic field gradient strength $\delta g$. The spin echo NMR signal is read out by an NV ensemble using the CASR pulse sequence, which consists of a train of single dynamic decoupling sequences. \textit{Insert:} A single dynamic decoupling subsequence, which consists of a train of $\pi$-pulses on the NV electronic spin, synchronized to the Larmor frequency of the nuclear spins. Typical parameter values used in this work are $\delta \approx 10$ ms, $\Delta \approx 80$ ms and $\tau \approx 75$ ms, whereas the gradient strength is swept from $g = 0$ $\mu$T$/$mm to $\sim$100 $ \mu$T/mm. }
\label{fig: Sequence}
\end{figure}

\textbf{Experimental setup.} The experimental setup developed for this publication is depicted in Fig. \ref{fig: Setup Schematic and grad} B, which can be split into two parts - the diffusion encoding using magnetic field gradient pulses during a spin echo sequence and the detection of the corresponding NMR signal with an NV ensemble.
We use a highly doped NV layer with a thickness of $\sim50 \,\mu$m that allows us to detect NMR signals on a similar length scale which also corresponds approximately to the typical diffusion displacement in our PGSE experiment \cite{Bruckmaier2021,Glenn2018}. As a model system/phantom, we use microfluidic chips \cite{Allert2022Micro}, where the NV-center layer forms the bottom wall of the microfluidic channel. The microfluidic chip is coupled to a syringe pump, allowing for precise control of the sample liquid. For the initialization and readout of the NV-center quantum state for the NMR detection, a 532 nm laser is coupled into the trapezoid diamond via a total internal reflection geometry\cite{Glenn2018}. This reduces laser-induced sample damage and heating while increasing the laser intensity at the NV layer\cite{Allert2022Micro}. A custom compound parabolic concentrator (CPC) is glued to the bottom side of the NV-diamond chip \cite{Wolf2015}. It efficiently collects the NV fluorescence which is then directed to a photodiode via a liquid light guide\cite{Arunkumar2022}. The NV diamond and microfluidic structure is imaged from the top, enabling us to correlate an optical image with the PGSE NV-NMR signal, defined by the location of the optical excitation. The free induction decay (FID) of the sample is induced by a radio frequency (RF) pulse and the corresponding NMR signal is detected via the NV ensemble, which is driven by microwave (MW) pulse sequence. The entire experiment is mounted within a large bore superconducting magnet, which provides a highly homogeneous and stable magnetic field ($B_0 \approx 0.175$ T), crucial for the detection of the NMR signal.\\
For the PGSE experiment, a set of three pairs of gradient coils ($\hat{x}$, $\hat{y}$ and $\hat{z}$) were designed and fabricated using the openly available gradient coil design tool CoilGen \cite{Amrein2022coilgen}. These coils have to satisfy unique conditions of NV-NMR spectroscopy, such as the optical access from multiple sides and, most importantly, a gradient along the $B_0$ field orientation, tilted at an angle of $\sim 54.74^\circ$ to the diamond surface normal. This angle is defined by the orientation of the NV centers within our diamond chip and, ultimately the crystal orientation of the diamond sensor \cite{Bruckmaier2021}. For our quantum sensing applications the external magnetic field B$_0$ is aligned along this NV axis, negating spin state mixing, which would otherwise strongly alter the NV-center spin dynamics \cite{Tetienne2012}. The method for finding optimal current carrying surfaces for this setup is described in Amrein \textit{et al.} \cite{Amrein2022}. Characterization was performed using ODMR of the NV centers in a widefield approach \cite{Levine2019}, extracting the relative $\hat{B}_0$ amplitudes over the diamond by measuring the NV-center Zeeman splitting, resulting in  experimentally assessed gradient sensitivities of $g_x \approx 29.74 \pm 0.09\, \frac{\mathrm{\mu T}}{{\mathrm{A\,mm}}}$, $g_y \approx 25.92\pm 0.09\, \frac{\mathrm{\mu T}}{{\mathrm{A\,mm}}}$ and $g_z \approx 23.27\pm 0.06\,\frac{\mathrm{\mu T}}{{\mathrm{A\,mm}}}$, respectively (Fig. \ref{fig: Sequence} A). In combination with the available current sources and under the constraints of air cooling in our proof-of-concept experiments we were able to reach gradient strengths of $100 \frac{\mathrm{mT}}{\mathrm{m}}$, which may appear rather weak to the standards of NMR microscopy, but which is on par with the top-performance whole-body clinical MRI scanners. \\

\textbf{Pulse sequence and theory.} For the NMR signal detection, we use the coherently averaged synchronized readout (CASR) method \cite{Glenn2018} (see Fig. \ref{fig: Sequence}B). It consists of a train of dynamic decoupling sequences which is synchronized to the sample FID. The detected signal of the optical NV readouts using CASR is an aliased version of the NMR signal. A more in-depth explanation of the sensing scheme is described in section 2 of the SM and in Glenn \textit{et al.}\cite{Glenn2018}. All experiments described in this work were conducted on protons in water, which were detected at a resonance frequency of $\sim$7.45 MHz (B$_0$ $\approx$ 0.175 T). To increase the NMR signal and reduce the averaging time, Overhauser dynamic nuclear hyperpolarization (DNP)  was used in all experiments by adding TEMPOL to our water sample \cite{Anet1965,Bucher2020}.  \\
The diffusion-NMR method used in this paper is called pulsed gradient spin echo (PGSE) \cite{Stejskal1965}. This sequence is a modification of the classic spin-echo experiment, where before and after the refocusing $\pi$-pulse two identical spatially varying $B_0$ gradient pulses are applied. The magnetic field gradient causes a spatially dependent Larmor frequency shifts which encodes the position of the nuclear sample spins. The first gradient pulse leads to a relative phase accumulation of each individual sample spin depending on its position, and the second gradient pulse leads to an inverse phase accumulation or refocusing up to the amount each spin has diffused along the gradient in the time between the two pulses. In the limit where the pulsed gradient amplitude is much higher than the constant background gradient of the magnetic field, the anisotropic diffusion coefficient ADC can be extracted by sweeping the strength of the applied gradient according to:

\begin{equation}\label{eq: PGSE diffusion}
    \mathrm{ADC} = -\ln(A/A_0)\, \left[  \left( \delta\,g\,\gamma  \right)^2 \,\left(\Delta-\frac{\delta}{3}  \right) \right]^{-1}\;,
\end{equation}

\noindent here $A$ and $A_0$ are the spin echo amplitudes with and without gradient pulses, respectively, $\delta$ is the duration, $\Delta$ is the spacing, $g$ is the strength of the applied gradient pulses and $\gamma$ is the gyromagnetic ratio of the sample spins\cite{Stejskal1965}. Fig. \ref{fig: Sequence} B and section 10 in the SM shows the corresponding pulse sequence and an experimental data set of PGSE NV-NMR experiment. The experiment is then repeated multiple times for averaging and a linear fit is performed to the log-scale of the resulting signal amplitudes to extract the ADC. The details are described in the material and methods and further in \textit{Kingsley et al.}\cite{Kingsley2006}.
For restricted diffusion, as is the case in our microfluidic channel, a slightly modified model including tensor properties of diffusion needs to be used to determine the apparent diffusion coefficient (ADC), which can be found in the SM, section 4. Individual tensor elements can be measured, by changing the direction of the first and/or second gradient pulse. The gradient directions used in this work are 1) parallel to $B_0$ ($\hat{z}$), 2) orthogonal to $B_0$ and parallel to the diamond surface ($\hat{x}$) and 3) the remaining direction at a $\sim$ 35.26$^\circ$ angle to the diamond surface normal ($\hat{y}$), as depicted in Fig. \ref{fig: Setup Schematic and grad} B.\\

\begin{figure}[H]
\centering
\includegraphics[width=\linewidth]{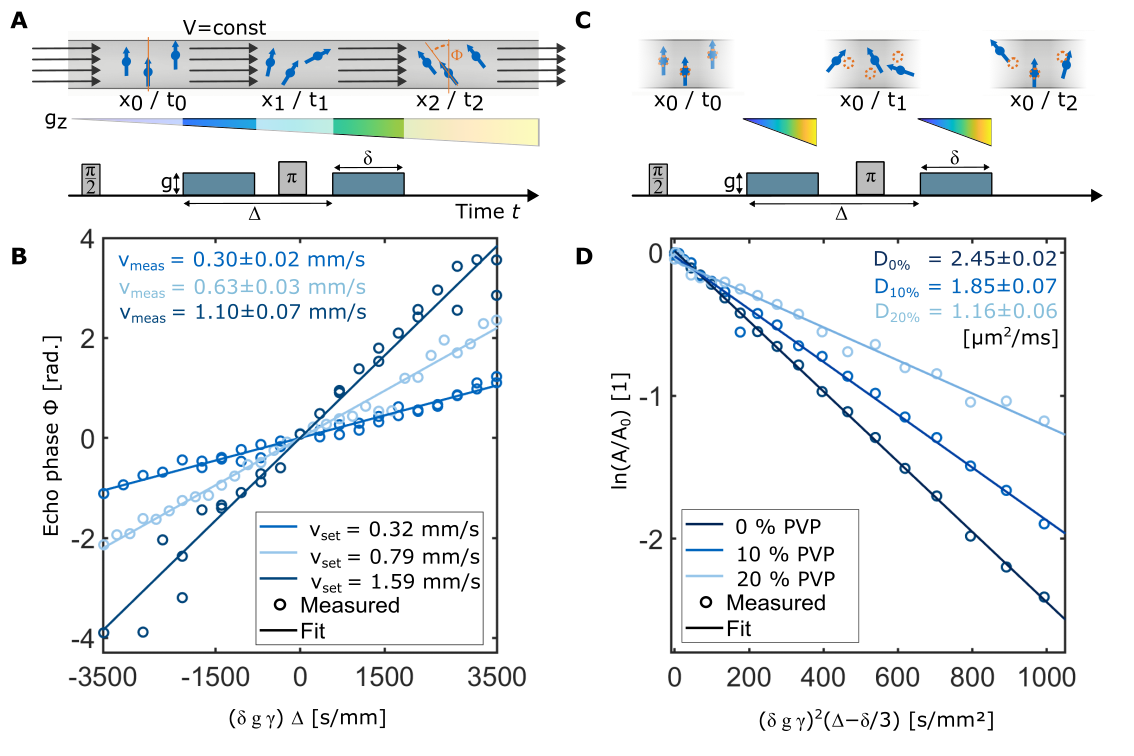}
\caption{\textbf{Measuring water flow and diffusion using PGSE NV-NMR.} \textbf{A)} Schematic of sample spins moving in a microfluidic channel during a velocimetry measurement. During the PGSE experiment a constant and laminar flow is applied which will lead to an equal translation of all sample molecules from their initial, $t = t_0$, location x$_0$ to their final position x$_2$ at $t = t_2$. This leads to an equal phase shift for each spin within the sample, which depends on the translated distance between the two gradient pulses and the strength and duration of the individual pulses. The PGSE-sequence is sketched in the bottom, the absolute gradient strength of each pulse is colour coded. \textbf{B)} Experimental data (phase $\phi$ as a function of gradient amplitude $g$) (circles) and fits (lines) for three different flow rates. Increasing the gradient amplitude leads to a larger phase accumulation due to the flow in the channel. \textbf{C)} Molecular diffusion leads to a random displacement of the sample spins which effectively attenuates the amplitude of the PGSE. The PGSE sequence is sketched in the bottom of the figure. \textbf{D)} Three PGSE measurements (sweeping the gradient strength) of different concentration of polyvinylpyrrolidone (PVP) K90 in water. The normalized spin echo amplitudes are displayed as circles and linear fits as dashed lines.}
\label{fig: PGSE-measurements}
\end{figure}

\textbf{Velocimetry measurements.} In the first set of experiments, we used our PGSE NV-NMR setup to measure the flow velocity of water within our microfluidic channel. Assuming a homogeneous flow profile, each molecule of water will have moved the same distance along the gradient during the free-diffusion time $\Delta$. This causes a common relative phase shift $\phi$ (Fig. \ref{fig: PGSE-measurements} A) of the nuclear spins. Since the NV-NMR detection method used for our experiments is phase sensitive \cite{Glenn2018}, the water flow within the channel can be calculated from the applied magnetic field gradient. Including laminar flow into equation \ref{eq: PGSE diffusion}, the combined effects of diffusion and translation on the sample magnetization can be described as \cite{Stejskal1965,Callaghan1991}:
\begin{equation}\label{eq: velocimetry}
    A/A_0 = \exp\left( i \;v\;\gamma \mathrm{g} \delta \;\Delta-  D\;(\gamma \mathrm{g} \delta)^2(\Delta-\delta/3) \right)\;,
\end{equation}
where $v$ is the flow velocity within the channel and $i$ is the imaginary unit. The signal phase $\phi = v \gamma g \delta \Delta$ can be extracted from the experimental data via the imaginary and real part of the spin echo's Fourier transformation. We would like to note, that conventional NMR methods exist, which directly measure the full propagator described in equation \ref{eq: velocimetry} \cite{Callaghan1988}. Plotting the phase $\phi$ against the magnetic field gradient strength allows us to determine the velocity from a linear fit \cite{Williamson2020}. Due to a complex interplay between the quadratic flow profile in our microfluidic channel and the homogeneous spatial sensitivity of NV-NMR spectroscopy \cite{Bruckmaier2021}, the recorded phase shift is not a simple linear relation to the set mean phase shift of the sample. The non-linearity was corrected by using numerical simulation, as described in SM section 5.

\begin{figure}[H]
\centering
\includegraphics[width=.9\linewidth]{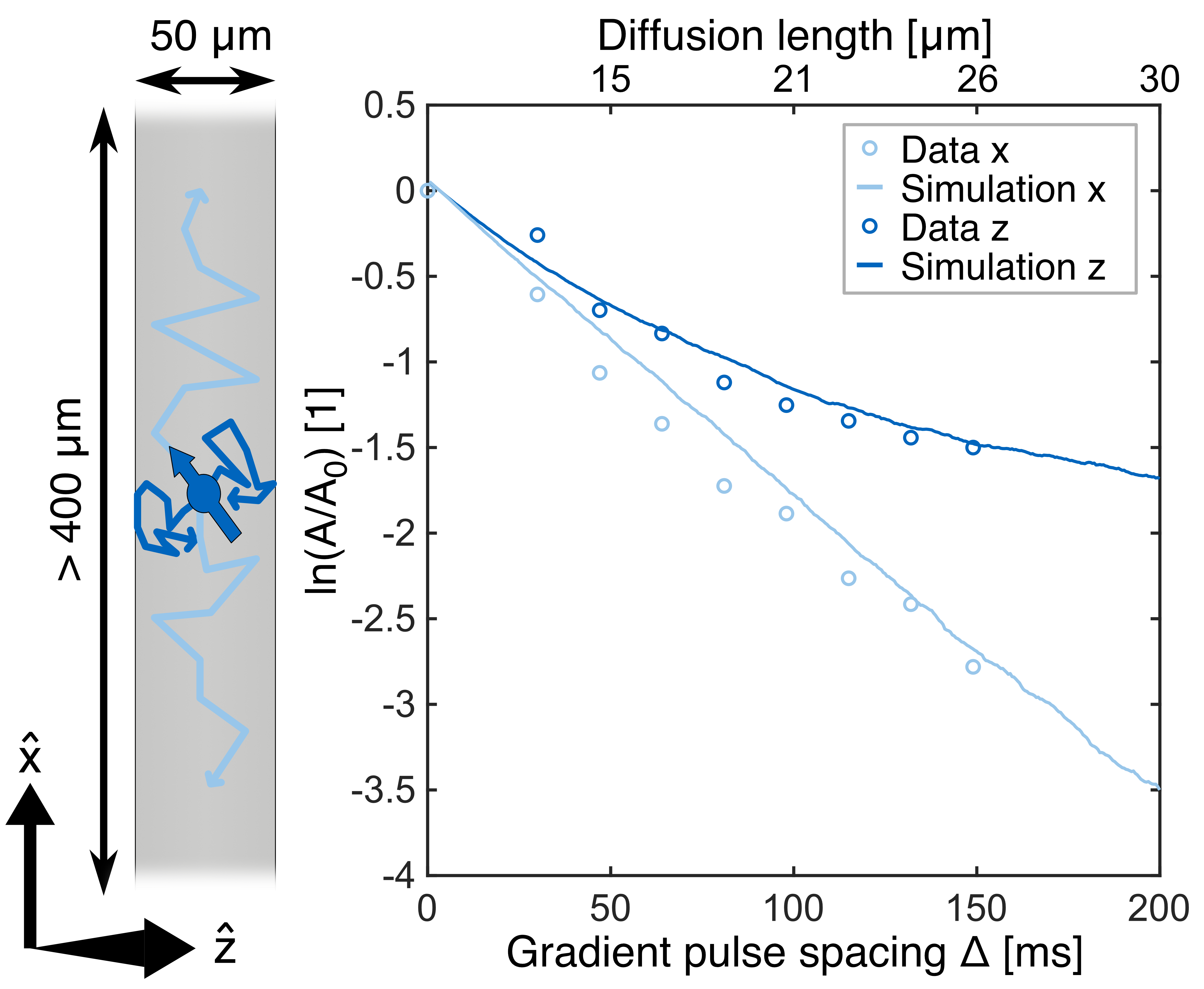}
\caption{\textbf{Investigating the time dependence of the ADC.} \textit{Left:} A sketch of the diffusion of our sample molecules in the microfluidic channel. The diffusion along the $\hat{z}$-direction is constrained, leading to an increased number of molecules interacting with the channel boundaries as the gradient spacing $\Delta$ is increased. In contrast, there are no effective restrictions in the x-direction. \textit{Right:} Simulation (line) and experimental (circles) data of a PGSE amplitude of the spin-echo signal is plotted on the y-axis. For reference the expected root-mean-squared distance diffused during $\Delta$ for the case of free diffusion is given on the top of the plot. Along the $\hat{x}$-direction (light blue) both simulation and experimental data predict an exponential decay of the spin-echo amplitude, as is the case for free diffusion, see equation \ref{eq: PGSE diffusion}. Along the $\hat{z}$-direction (dark blue), the diffusion is restricted by the channel walls, and an increasing number of molecules interact with the channel walls and the ADC decreases over time. This leads to a slower decay of the spin-echo amplitude.}
\label{fig: delta sweep}
\end{figure}

\begin{figure}[H]
\centering
\includegraphics[width=.9\linewidth]{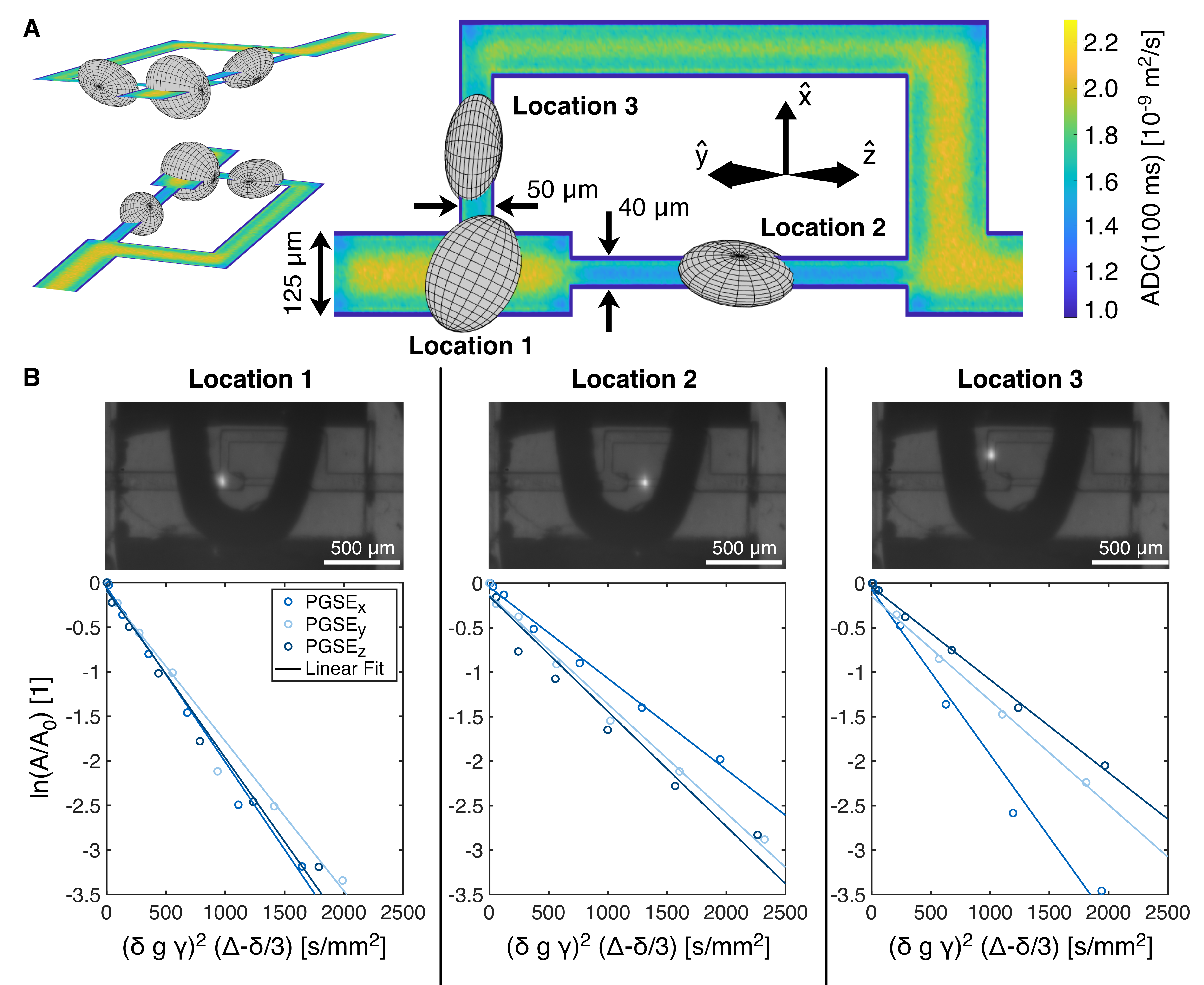}
\caption{\textbf{ Spatially resolved PGSE NV-NMR experiments within microfluidic structures.} \textbf{A)} Visualization of the measured diffusion tensors in our microfluidic channel. The size of the ellipsoid corresponds to the strength of the ADC in this direction. Three different perspective of the ellipsoids are displayed for a better visualization. The color of the microfluidic channel corresponds to the trace of the numerically simulated diffusion tensor after $\Delta = 100$ ms of free diffusion. \textbf{B)} Photographs of the investigated locations (top) taken with a camera. Exemplary data sets of each of the three locations (bottom), in location 1 the diffusion is close to free diffusion for all directions, while in location 2 and 3 the $\hat{x}$ and $\hat{y}$ or $\hat{z}$ directions are more restricted, respectively. } 
\label{fig: spatial-resolution}
\end{figure}

\begin{table}[H]
    \begin{center}
    \caption{\textbf{Results of the PVP diffusion measurements.} Literature\cite{Mills1973,Harris1980,Tofts2000}, simulated and measured values of the ADC for three different concentrations of PVP in water (w/w) with their respective uncertainties. }
\begin{tabular}{|l|l|l|l|}
\hline
ADC$_z$ $[\mu m^2/ms]$  at $\sim $25 $^\circ$C    & 0\% PVP  & 10\% PVP & 20\% PVP \\ \hline
Literature                & 2.31      & 1.81      & 1.37      \\ \hline
Simulation                & 2.14      & 1.69      & 1.13      \\ \hline
Experimental result       & 2.45$\pm$0.02 & 1.85$\pm$0.07 & 1.16$\pm$0.06 \\ \hline
\end{tabular}
    \end{center}
  
\end{table}\label{tab: D values}
 The experiments were conducted in a straight microfluidic channel with dimensions of 80 $\mu$m (orthogonal to the diamond surface) x 100 $\mu$m (along the $\hat{x}$-direction) x 2000 $\mu$m (along the diamond surface) \cite{Allert2022Micro}. The experimentally measured flow rates by PGSE NV-NMR were  $0.30\pm0.02\;$mm/s, $0.63\pm 0.03\;$mm/s and $1.10\pm0.07\;$mm/s, which are slightly but consistently lower than the parameters set at the syringe pump (Table 1 and Fig. \ref{fig: PGSE-measurements} B). This can be explained by an additional layer of glue in between the diamond and the microfluidic chip, increasing the effective volume of the channel: The flow rate in the microfluidic channel is calculated from the flow rate set at our pump, given in units of $\frac{volume}{time}$. This is divided by the intended cross section of the microfluidic channel, resulting in the flow rates $v_{set}$ as seen above. Any difference between the cross section of the microfluidic channel in the experiment and as designed would lead to a proportional offset between measured and set flow rates. \\
\textbf{Diffusion measurements.} In the second set of experiments, we measured the diffusion coefficient of water in the microfluidic channel. In contrast to the laminar flow in the previous experiment, diffusion leads to random motion and a reduction of the spin echo amplitude as a function of gradient strength (Fig. \ref{fig: PGSE-measurements} C). We used water doped with varying concentrations of an organic polymer polyvinylpyrrolidone (PVP) K90 (0\%, 10\% and 20\% w/w) at $T\approx$ 25 $^\circ$C, to modify the diffusivity of water, similar to previous reports \cite{Mills1973,Harris1980,Tofts2000,Wagner2017}. Since we measure the water diffusion within a microfluidic channel, the free diffusion will be attenuated by its boundaries. For that reason, we chose to sweep the amplitude of the $g_z$ gradient to measure the ADC, since the diffusion along this direction is the least restricted therefore the closest to the values reported in the literature. Nevertheless, the boundaries of the microfluidic channel will reduce the ADC's diagonal elements, compared to the free diffusion case. Therefore, we simulated the expected ADC based on the literature values as described in the SM, section 4. The resulting data can be found in Table 1 and seen in Fig. \ref{fig: PGSE-measurements} D. The expected and simulated values are in good agreement with the values obtained from experimental PGSE NV-NMR. The remaining discrepancy between measured and simulated parameters can be explained by possible sample heating, as discussed in the SM, section 7, which affects the diffusion in solutions with higher concentrations of PVP to a lesser degree \cite{Wagner2017}. \\
\textbf{Investigating the time dependence of ADC.} Having established the ability to perform PGSE experiments in combination with NV-NMR, we investigate the effects of the restricted diffusion in one of our microfluidic channels. The one-dimensional $ADC$ can be defined as:
\begin{equation}
    ADC = \frac{\braket{x^2}}{2 \Delta}\;,
\end{equation}
\noindent  where $\Delta$ is the free diffusion time during our PGSE experiment. In the case of free diffusion, this relationship is constant and $ADC$ is independent of $\Delta$. In the case of restricted diffusion though, the root-mean-squared distance diffused $\braket{x^2}$ is limited by the length scale of the restriction. As the free diffusion time $\Delta$ increases, more and more molecules will interact with the confinement boundaries and the $ADC$ will tend to $0$. To investigate this phenomenon, we performed experiments in a microfluidic channel with strong confinements along the direction perpendicular to the channel and effectively no confinements along the longitudinal direction. The experimental results were verified by numerical simulations, both data are shown in Fig. \ref{fig: delta sweep}. An image of the corresponding microfluidic channel can be found in Fig. \ref{fig: spatial-resolution}. The simulations are described in detail in the Materials and Methods section as well as in the SM section 4. Along the longitudinal direction we expected to find an unrestricted diffusion of the sample molecules, leading to linear dependence of the spin-echo amplitude on the free diffusion time in the logarithmic scale (equation \ref{eq: PGSE diffusion}). This is verified by both experimental data and the numerical simulations. Along the perpendicular direction we expect a slower signal decay, since the root-mean-squared distance diffused along this direction is limited and, as described above, the ADC will decrease with increasing $\Delta$. This is clearly evident in both the experimental data and the conducted simulations.\\
\textbf{Measuring spatially resolved anisotropic diffusion.} Having demonstrated the effect the microfluidic channel walls have on the free diffusion of our sample molecules, we continue to probe water diffusion spatially resolved within microstructures. For that purpose, we designed and fabricated a microfluidic structure with different channel sizes and orientations. Due to the optical readout of the PGSE NV-NMR signal, any location within this structure can be probed by moving the NV-excitation laser (Fig. \ref{fig: spatial-resolution} A and B). Our current setup has a limited field of view of approximately 1 mm² and requires scanning with the laser point-by-point for spatially-resolved measurements. However, future developments may employ wide-field imaging to alleviate this issue \cite{Levine2019}.

 For an estimation of the expected ADC within our microfluidic structure, we simulated the ADC for each point using particles undergoing a random walk. Due to the small length scales, the ADC along the different cardinal directions can vary drastically (Fig. \ref{fig: spatial-resolution}). Then, we performed PGSE NV-NMR experiments at three different locations and along six different directions within this structure. From these six, non-colinear measurements the diffusion tensor is calculated according to Kingsley \textit{et al.}\cite{Kingsley2006}. The resulting tensors are depicted in Fig. \ref{fig: spatial-resolution} A. The ADC changes depending on the channel dimensions at the location of the laser spot within the structure, in accordance with our simulation (Fig. \ref{fig: spatial-resolution} C). The difference is most pronounced when measurement locations 2 and 3 are compared. In location 2, the microfluidic channel has a cross section of 40x40 $\mu$m$^2$, meaning the diffusion is strongly constrained ($\sim$ 40 $\mu$m) in  and $\hat{y}$ direction, while it is close to free along the $\hat{z}$ direction. On the other hand in location 3, the channel has a cross section of 50 x 80 $\mu$m$^2$, the longitudinal direction can be considered totally free diffusion and the shorpet perpendicular direction is constrained within ($\sim$ 50 $\mu$m). We expect the eigenvectors of the ADC tensor to be orthogonal to the walls of the approximately cuboid microfluidic channel. The tensors calculated from our measurements, seen in Fig. \ref{fig: spatial-resolution}, are close to this expected orientation, as described in SM section 8, and the resulting ADC is in good agreement with the simulation. Simulated and experimental results for each of the locations can be found in table 2.

\begin{table}[H]
\caption{\textbf{Results of the spatially resolved PGSE NV-NMR experiments within the microfluidic structure.} Measured and simulated elements of the measured ADC tensor for each of the locations investigated. An exemplary data set and the full tensor including the off-diagonal elements can be found in the SM section 10.}
    \begin{center}
        \begin{tabular}{|l|l|l|l|l|}
\hline
\multicolumn{2}{|c|}{ADC $[\mu m^2/ms] $ at $\sim$25$^\circ$ C}     & Location 1        & Location 2        & Location 3         \\ \hline
\multirow{2}{*}{ADC$_x$} & simulated        & 2.19 $\pm$ 0.10& 1.06 $\pm$ 0.10& 2.29 $\pm$ 0.10 \\ 
                       & measured         & 2.24 $\pm$ 0.13 & 1.19 $\pm$ 0.13 & 2.24 $\pm$ 0.07 \\ \hline
\multirow{2}{*}{ADC$_y$} & simulated        & 1.92 $\pm$ 0.10& 1.75 $\pm$ 0.10& 1.36 $\pm$ 0.10 \\ 
                      & measured         & 1.93 $\pm$ 0.16 & 1.67 $\pm$ 0.32 & 1.39 $\pm$ 0.29 \\ \hline
\multirow{2}{*}{ADC$_z$} & simulated        & 2.01 $\pm$ 0.10 & 2.03 $\pm$ 0.10& 1.16 $\pm$ 0.10\\ 
                       & measured         & 2.16 $\pm$ 0.15 & 2.10 $\pm$ 0.31 & 1.18 $\pm$ 0.05 \\ \hline
\end{tabular}
  \end{center}
\end{table}\label{tab: D values2}

\section*{Discussion}
In this work we have demonstrated spatially resolved PGSE experiments within microstructures using NV centers in diamond. We would like to note that the spatial resolution has not yet reached any physical limitations. Higher spatial resolution  can be achieved by decreasing the thickness of the diamond's NV-center doped layer and reducing the diameter of the excitation laser beam. In our current experimental setup, both the NV-center doped layer and the diameter of the laser location are on the order of $\sim 50\;\mu$m, limiting our spatial resolution to the same order of magnitude \cite{ Bruckmaier2021}. Reaching the optical diffraction limit is feasible, although with the drawback of highly reduced sensitivity which would lead to long averaging times, as discussed in the SM, section 6. Nevertheless, this technique could enable quantifying the diffusion properties of basic micro-structural building blocks on the single cell level, which would help to validate current models in medical MRI \cite{Fieremans2018,Novikov2018}. However, for biological applications Overhauser DNP used in our study is not recommended, due to the need for polarizing agents within the sample and strong MW fields. Alternatives are increasing the magnetic field $B_0$ strength (\textit{e.g.} to 1 T) or other, bio-compatible hyperpolarization methods \cite{Eills2022} (for instance, dissolution DNP (which also requires a polarizing agent) or parahydrogen induced hyperpolarization, PHIP \cite{Arunkumar2021}). The latter methods allow for high signal enhancements by relying on promptly injected hyperpolarised agents and in combination with NV-NMR would potentially allow for probing the diffusion of metabolites in single cells.\\
Another unique feature of our method is the possibility of applying very strong magnetic field gradients  \cite{Kimmich1991,Callaghan1998}. Due to the small length scale of NV-NMR gradient coils can be miniaturized, and thereby can achieve up to 10 mT/$\mu$m \cite{Zhang2017,Arai2015}. For such extremely high gradient fields, concomitant field components that occur as an unavoidable consequence of Maxwell's equations, may require corrective actions either by designing specialized compensated pulse sequences \cite{Szczepankiewicz2019} or by increasing the main magnetic field strength $B_0$. These technical developments may provide unique insights, \textit{e.g.} in detecting slowly diffusing spins \cite{Heenen2017,Benedek2020} (such as Li ions in solid state materials), in detecting diffusion on smallest length scales, or elucidating the origin of “dot-compartments”, small diffusion-restricted spaces in tissues, which are currently discussed in literature \cite{Tax2020}.\\

In summary, we have developed a powerful NV-based NMR method, which enables us to image diffusion on microscales. The technique allows for the local detection of water flow and diffusion
within microscopic sample volumes. Finally, we demonstrated the capability to measure the ADC spatially resolved within a model microstructure in three directions, which showed restriction in diffusion due to the local geometry. Although our current spatial resolution is comparable with published conventional diffusion MRI results \cite{Jelescu2014,Johnson2023,Wu2013}, we are limited by the optical resolution rather than the spatial encoding with magnetic field gradients \cite{Callaghan1988}. We anticipate that further technical improvements will allow us to approach a spatial resolution of a single micrometer and thereby shed new light on microstructure diffusion by decoupling the long-standing link between spatial encoding and diffusion weighting in NMR. Our technique and experiments mark a major milestone towards probing single cells, tissue microstructures or ion-conducting materials in energy research. 


\section*{Materials and Methods}

\textbf{Experimental setup.}
A schematic of the experimental setup and its optics is shown in the SM, section 3, Fig. S4 and S5. An electronic-grade, single crystal, 100-oriented diamond (2x2x0.5 mm, Element Six, Oxford, UK) which has been overgrown with a $\sim$ 19 ppm nitrogen-doped $^{12}$C and $^{15}$N isotopically enriched diamond layer with a thickness of $\sim$ 50 $\mu m$ by the Fraunhofer Institute for Applied Solid State Physics (Freiburg, Germany) as described in Sch\"{a}tzle \textit{et al.}\cite{Schtzle2022} and cut into a trapezoidal shape, which was then electron irradiated and annealed to increase the nitrogen to NV-conversion rate. This particular thickness of the nitrogen doped layer was chosen, since simulations of the experimental geometry indicated this to optimize the signal to noise ratio (SNR) for our microfluidic channels \cite{Allert2022Micro}. Ramsey and Hahn echo spectroscopy is used to measure an NV-ensemble $T_{2}^{*}$ dephasing time of $\sim$0.65 $\mu$s and $T_{2}$ decoherence time of $\sim$ 9 $\mu$s, respectively. The diamond was glued into a microfluidic chip designed in house and fabricated by LightFab GmbH (Aachen, Germany) using Norland Optical Adhesive 68 UV-curing glue \cite{Allert2022Micro}. The assembled microfluidic chip with diamond was positioned in a custom build superconducting magnet (3T-215-RT, Superconducting Systems INC., Billerica, USA) and one of the four possible NV orientations with the diamond lattice was aligned with the external magnetic field ($B_0\approx 0.175$ T). Flow pumps (AL-1000, World Precision Instruments, Sarasota, USA) were used to control the flow velocity within the microfluidic channel. The diamond's fluorescent light was collected using a compound parabolic concentrator (CPC) glued to the bottom of the microfluidic chip (designed in house and fabricated by S\"ud-Optik Schirmer GmbH, Kaufbeuren, Germany). The CPC output was attached to a custom made liquid light guide (Lumatec GmbH, Munich, Germany) which directs the fluorescence outside of the magnet through a long pass filter (BLP01-647R-25, Edge Basic 647 Long Wave Pass, Semrock, Rochester, USA) onto a balanced photodiode (PDB210A, Thorlabs, Bergkirchen, Germany). A reference laser beam was used for efficient laser noise cancellation. The photodiode's voltage was read out using a data acquisition unit (NI USB-6821, National Instruments, Austin, USA).\\
The NV-center spins were initialized using a $532 \; nm$ laser (Laser Quantum Opus 53, Novanta Photonics, Wackersdorf, Germany) with a power of about $\sim$ $380$ mW. Initially the laser passes an opto-acoustic modulator (3260-220, Gooch and Housego, Ilminster, UK) to generate pulses of a typical length of $5 \; \mu s$. A multi-order half-wave plate (WPMH05M-532, Thorlabs, Bergkirchen, Germany) was used to adjust the polarization of the laser light for efficient NV excitation. Finally, the laser beam was expanded (BE02-05-A, Thorlabs, Bergkirchen, Germany) and focused onto the diamond using a $f = 250$ mm lens (LA1433-B-ML, Thorlabs, Bergkirchen, Germany), resulting in a beam diameter of $1/e^2\approx 45 \,\mu$m FWHM. The position of the laser location within the microstructure was imaged from the top on a camera (a2A3840-45umBAS, Basler, Ahrensburg, Germany). \\
The whole experimental sequence was controlled by an arbitrary waveform generator (AWG70000B, Tektronix, Beaverton, USA). It synchronizes all other devices (signal sources, switches, data acquisition unit and opto-acoustic modulator) via synchronised transistor transistor logic (TTL) signals (SM, section 3 Fig. S5). The pulse sequence, driving the NV-center spins, is programmed and uploaded with a 500 MHz carrier frequency and up-converted using an IQ mixer (mmiq0218LXPC, Marki Microwave, Morgan Hill, USA) and a MW signal source (SMB100A, Rhode und Schwarz, Munich, Germany). The resulting MW pulses were then amplified using a broadband $50 \; W$ amplifier (AMP1016, Exodus, Las Vegas, USA) and delivered using a home-built microwave antenna \cite{bucher_quantum_2019}. A RF source (LXI DG1022, Rigol, Suzhou, China) was amplified (LZY-22+, Mini-Circuits, Brooklyn, USA) and connected to two coils in a Helmholtz geometry with radius $R = 1.5$ cm for driving the sample nuclear spins with Rabi frequencies up to $6.3 $ kHz. An additional coil for calibration purposes was connected to another RF source (LXI DG1022, Rigol, Suzhou, China), to determine the sensitivity of our experiment as described in Glenn \textit{et al.} \cite{Glenn2018}. A third RF source (LXI DG1022, Rigol, Suzhou, China) was used to generate the gradient pulses which were fed into a bipolar power supply (BOP 5-20DL, Kepco, Naju, South Korea) capable of $\pm \; 20\;V$ and $\pm \; 5\;A$, which in turn was connected to the gradient coils (Beta-Layout, Aarbergen, Germany). The microfluidic chip, MW, RF and gradient coils were all mounted on a custom designed, 3D-printed sample holder (grey v4 resin, Form 3, Formlabs, Somerville, USA). A photo of the setup and the assembly is depicted in the SM, section 3, Fig. S6. 
To achieve a stronger NMR signal, Overhauser DNP was used\cite{Bucher2020}. In all experiments a 10 mM concentration of 4-Hydroxy-2,2,6,6-tetramethylpiperidin-1-oxyl (TEMPOL, 581500, Sigma-Aldrich, St. Louis, USA) was added to the respective sample. TEMPOL is a stable radical, which under continuous, strong and resonant MW radiation ($0.3\,\mathrm{s}$) can hyperpolarize the nuclear sample spins, leading to a $\sim 200$-fold increase in the NMR signal strength\cite{Bucher2020}.

    \textbf{Chemicals.}
    The polyvinylpyrrolidone K90 (PVP, 81440, Sigma-Aldrich, St. Louis, USA) and the 4-Hydroxy-2,2,6,6-tetramethylpiperidin-1-oxyl (TEMPOL, 581500, Sigma-Aldrich, St. Louis, USA) were purchased from Sigma-Aldrich and used without further purification steps; the chemical structures are depicted in the SM, section 9. The solutions were prepared using deionized water with a resistivity of 18.2 M$\Omega$ cm (MilliporeSigma, Burlington, United States).

\textbf{Gradient coil design.}
A Matlab-based software package\cite{Amrein2022coilgen} based on the stream function method \cite{lemdiasov_stream_2005} was used for the design of the gradient system. A biplanar configuration was chosen for the geometry of the gradient coils since it allows for better access for the fluorescence optical readout path compared to other geometries. Searching for a suitable biplanar configuration, several geometrical parameters were investigated such as plate size, plate distance and plate orientation and after evaluation, a solution with a plate size of 50 mm, a plate distance of 30 mm and an atypical azimuth plate tilt of $\sim 35.26^\circ$ against the $B_0$ magnetic field was selected for printed circuit board (PCB) fabrication. Although the value of $\sim 35.26^{\circ}$ for the azimuthal inclination is not optimal for the gradient's strength (the optimum is found at $55^{\circ}$), the gradient plates mounted vertically present a reasonable compromise between the achievable performance and the compatibility with the NV-NMR experimental setup. Since the gradient coils are only added for diffusion weighting, thermal limitations are not expected if the duty cycles of the used MR sequence are sufficiently low. More information on the design can be found in the dedicated publication by Amrein \textit{et al.} \cite{Amrein2022}.

\textbf{Microscale NV-NMR using the CASR pulse sequence.}
 For this work the universally robust dynamic decoupling sequence (UDD) \cite{Genov2017} containing 12 pulses is used, with typically 50 repetitions per $\pi$-pulse train, leading to $\approx$ 600 $\pi$-pulses per measurement step. A typical duration of our $\pi$-pulses is $\sim 30$ ns. In the case of the CASR pulse sequence, a detuning $\delta f$ to the peak frequency $f_0$ is detected typically in the range of $|\delta f| < 3000$ Hz \cite{Glenn2018}. The typical AC sensitivity and volume normalized AC sensitivity of our experiment were $\sim20 $ pT$/\sqrt{\mathrm{Hz}}$ and $\sim 5.6$ nT$\,\sqrt{\mu\mathrm{m}^3}/(\sqrt{Hz})$, respectively. An example of a CASR measurement using the universally robust dynamic decoupling-8 sequence is depicted in Fig. \ref{fig: Sequence}. More information on the CASR method can be found in Glenn \textit{et al.}\cite{Glenn2018}.
 
\textbf{PGSE NV-NMR pulse sequence.} 
 Typical parameter values used in for the PGSE sequence are $\delta \approx 10$ ms, $\Delta \approx 80$ ms and $\tau \approx 75$ ms, sweeping the gradient strength from $g \approx 0$ mT$/$m to $100$ mT$/$m. Experiments were averaged 100 times each, usually waiting a total of 3 s in between averages, to allow relaxation of the sample nuclear spins to thermal equilibrium. The typical single-shot SNR of a hyperpolarized water NMR signal in our experiments was $\sim 100$. Typical coherence times of the water protons were $T_2* \approx 60$ ms, $T_2 \approx 80$ ms and $T_1 \approx 300$ ms. $T_2*$ is very likely limited by magnetic field inhomogeneities of our experimental setup, while $T_2$ and $T_1$ were limited through the addition of TEMPOL into the solution.  \\

\textbf{Wide-field gradient imaging using CW-ODMR.}
The magnetic field gradients, shown in Fig. \ref{fig: Setup Schematic and grad} C, are measured by wide-field DC magnetic imaging using continuous-wave optically detected magnetic resonance (CW-ODMR) \cite{Levine2019}. As sensor we use an electronic-grade diamond chip (1.9 mm x 1.9 mm x 0.5 mm) with a 14 $\mu$m thick, $^{12}$C and $^{15}$N isotopically enriched, nitrogen doped layer (nitrogen concentration $\sim$ 2.3 ppm), which was electron irradiated and annealed to increase the nitrogen to NV-center conversion rate. An external magnetic field $B_{0}$ of $\sim$ 4.4 mT is applied along the NV-symmetry axis which lifts the degeneracy of the $\mathrm{m_s}$ = $\pm$ 1 states. For excitation green laser light (Saphire LPX, Coherent, 	Santa Clara, USA) is used to fully illuminate the diamond chip ($\sim$ 600 mW). Electron spin ground state $\mathrm{m_s}$ = 0 $\rightarrow$ $\pm$1 transitions are probed by sweeping an applied microwave field in 400 steps (200 steps for each transition) synchronized to the NV-fluorescence readout. The NV-spin driving field is produced by a signal source (SMB100A, Rhode und Schwarz, Munich, Germany), amplified (ZHL-16W-72+, Mini-Circuits, Brooklyn, USA) and delivered to the diamond sample by a MW antenna. The NV fluorescence was passed through a spectral filter (BLP01-647R-25, Edge Basic 647 Long Wave Pass, Semrock, Rochester, USA) and collected by a camera (a2A1920-160umBAS, Basler, Ahrensburg, Germany) with a magnification of 2.75. For the measurements, 4x4 pixels are binned on the camera, resulting in 480x304 data points. Each data point is recorded with an exposure time of 600 $\mu$s and 800 averages. Thus, we acquire an image stack with a depth of 400, where each pixel stack corresponds to a single CW-ODMR spectrum. Four different CW-ODMR spectra are recorded with and without (background) applying a current of one ampere to the $\hat{x}$, $\hat{y}$ and $\hat{z}$ gradient coils. The gradient fields along $B_0$ are obtained by fitting (double Lorentzian) the NV-resonance lines of the collected data after subtraction the $B_{0}$ background field.  Magnetic field values are calculated for each pixel stack from the splitting of the $\mathrm{m_s}$ = $\pm$ 1 states ($2 \gamma B_{0}$), resulting in a 2D magnetic field map. The fitted gradients in g$_y$  and g$_z$ direction were corrected using factors of $\frac{1}{\sin(35.26^\circ)}$ and $\frac{1}{\cos(35.26^\circ)}$ respectively, since the gradient direction is not parallel to the diamond surface. Any constant offset produced by the gradient coils can be neglected, since it will have a negligible effect on the echo amplitude of the sample magnetization.

\textbf{Data analysis.}
The PGSE experiments were typically averaged 100 times and the data from the end of the second magnetic field gradient pulse to the point in time, where the FID's spectral components where below the noise floor, usually after about 250-350 ms, was used for analysis. This window was zero filled to a total of three times the initial length and Fourier-transformed. The NMR signal peak was integrated and the resulting data was normalized to the data point with the highest signal amplitude, taking into account the possibility of constant background gradients. Finally the whole data set was fitted with the function G:
\begin{equation}
    G(D, g, \delta g) = \ln(\exp\left(-D \, (\delta \gamma (g-\delta g))^2\,(\Delta -\delta/3)\right)+\mathrm{Offset})\;.
\end{equation}
$\Delta$, $\delta$ and $\gamma$ are known, the gradient amplitude $g$ is swept. $\delta g$ is a fit parameter which takes constant magnetic field inhomogeneities caused by magnetic susceptibility mismatches between sample, microfluidic and diamond chip into account. For the calculation of the ADC tensor, we followed Kingsley \textit{et al.}\cite{Kingsley2006}.

\textbf{Simulation of diffusion in a restricted volume.}
The simulations done for this work are based on a random walk of individual sample particles, in a defined, micro-scale volume. A similar technique is used in Cartlidge \textit{et al.}\cite{Cartlidge2022} to simulate the diffusion induced relaxation in porous media. In each iteration a Gaussian distributed, random distance in an equally distributed random direction is chosen and the particle is moved accordingly. If the path hits a boundary of the micro-scale bounding volume, the packet is reflected inwards the rest of the way. At each time step the root mean squared distance traveled can be calculated, which is directly related to the ADC. For more information see the SM, section 4 and Bruckmaier \textit{et al.}\cite{Bruckmaier2021}.

\section*{Acknowledgements}

This project was funded by the European Research Council (ERC) under the European Union’s Horizon 2020 research and innovation program (grant agreement No 948049) and by the Deutsche Forschungsgemeinschaft (DFG, German Research Foundation) - 412351169 within the Emmy Noether program. D.B.B. acknowledges support from the DFG under Germany’s Excellence Strategy—EXC 2089/1—390776260 and the EXC-2111 390814868.\\
P.A., S.L. and M.Z. acknowledge support of the German Federal Ministry of Education and Research (BMBF) grant number 13GW0356B.

\subsection{Data and Materials Availability}
All data needed to evaluate the conclusions in the paper are present in the paper and/or the Supplementary Materials. 

\section*{Author contributions statement}
D.B.B. conceived and supervised the study. F.B. designed the experiment and performed all simulations and analytic derivations. F.B. and R.D.A. conducted the experiments. R.D.A. and N.N. contributed to various parts of the experimental setup. R.D.A. designed the microfluidic chips with help from N.N. and F.B.. K.D.B. designed and N.N. conducted the wide-field ESR experiments. P.A., S.L. and M.Z. designed the gradient coils used in this experiment and helped with their characterization. P.K., P.S. and R.D.A. produced the diamonds used in the experiments. F.B.,  R.D.A and D.B.B. analyzed and discussed the data and wrote the manuscript with inputs from all authors.

\section*{Additional information}

All authors declare that they have no competing interests.\\
Additional information on this work can be found in the Supplementary Material.

\section*{Bibliography}

\nocite{*}
\bibliographystyle{unsrt}
\bibliography{main}

\flushbottom

\end{document}